\documentclass[useAMS,usenatbib]{mn2e}

\usepackage[psamsfonts]{amssymb}
\usepackage[dvips]{graphicx}
\usepackage{graphicx}
\usepackage{amsmath,alltt}                                                                                                                          
\usepackage{multirow}
\usepackage{rotating}
\usepackage{lscape}
\usepackage{tablefootnote}
\usepackage{hyperref}
\title[Central Star Clusters in Dwarf Galaxies]{The Structural and Kinematic Evolution of Central Star Clusters in Dwarf Galaxies and Their Dependence on Dark Matter Halo Profiles}
\author[Webb $\&$ Vesperini]{Jeremy J. Webb $^{1,2}$ \& Enrico Vesperini $^2$
\thanks{E-mail: webb@astro.utoronto.ca (JW), evesperi@indiana.edu (EV)} \\
$^{1}$ Department of Astronomy and Astrophysics, University of Toronto, 50 St. George Street, Toronto ON M5S 3H4, Canada \\
$^{2}$ Department of Astronomy, Indiana University, Swain West, 727 E. 3rd Street, IN 47405 Bloomington, USA}

\begin{document}

\pagerange{\pageref{firstpage}--\pageref{lastpage}} \pubyear{2015}

\maketitle

\label{firstpage}

\begin{abstract}

Through a suite of direct $N$-body simulations, we explore how the structural and kinematic evolution of a star cluster located at the center of a dwarf galaxy is affected by the shape of its host's dark matter density profile. The stronger central tidal fields of cuspier halos minimize the cluster's ability to expand in response to mass loss due to stellar evolution during its early evolutionary stages and during its subsequent long-term evolution driven by two-body relaxation. Hence clusters evolving in cuspier dark matter halos are characterized by more compact sizes, higher velocity dispersions and remain approximately isotropic at all clustercentric distances. Conversely, clusters in cored halos can expand more and develop a velocity distribution profile that becomes increasingly radially anisotropic at larger clustercentric distances. Finally, the larger velocity dispersion of clusters evolving in cuspier dark matter profiles results in them having longer relaxation times. Hence clusters in cuspy galaxies relax at a slower rate and, consequently, they are both less mass segregated and farther from complete energy equipartition than cluster's in cored galaxies. Application of this work to observations allows for star clusters to be used as tools to measure the distribution of dark matter in dwarf galaxies and to distinguish isolated star clusters from ultra-faint dwarf galaxies.

\end{abstract}


\section{Introduction} \label{intro}

The ability to measure how dark matter (DM) is distributed across large and small scales allows for constraints to be placed on the various cosmological models used to describe the Universe. A key prediction of the $\Lambda$ Cold Dark Matter ($\Lambda CDM$) model, one of the most widely used and tested paradigms, is that DM exists in the form of halos that can range both in size and mass between galaxy cluster halos to dwarf galaxy halos. Over large scales, $\Lambda CDM$ has been successful in reproducing the distribution of DM halos \citep[e.g.][]{wright92, tegmark02, bennett13, planck14, anderson14}. 

Over small scales, comparable to that of a galaxy or dwarf galaxy, $\Lambda CDM$ predicts that the inner DM density profile will be cusp-like in shape. More specifically, the density of DM $\rho$ will steeply rise as distance from the halo centre $r$ decreases ($\rho \propto r^{-1}$) \citep{dubinski91, navarro96}. Conversely, dynamical models of dwarf galaxies and the observed inner rotation curves of galaxies indicate the presence of a DM core ($\rho \propto r^{0}$) \citep{moore94,flores94,battaglia08,walker11,amorisco12,agnello12,adams14,oh15}. This discrepancy represents the well known core-cusp problem, with a number of studies suggesting that the effects of baryonic feedback represent a potential remedy (see \citet{bullock17} for a recent review and references therein). Other investigations have explored alternative DM models, as warm DM, self-interacting DM, and 'fuzzy' DM have all been shown to produce more core-like halo density profiles than $\Lambda CDM$ \citep[e.g.][]{press90, hu00, spergel00, vogelsberger12, elbert15, ludlow16, hui17}. Alternatively, \citet{genina17} showed that deviations from the spherical symmetry assumption made in the calculations of mass estimates might be the main issue behind the core-cusp problem. Hence proper measurements of DM halo shapes are necessary to constrain and rule out different DM models. 

Estimates of the size and shape of the Milky Way's DM halo have been done using a variety of methods, most recently using the proper motions of stellar streams \citep[e.g.][]{bonaca14, bovy16, sanderson17} and star clusters (SCs) \citep[e.g.][]{eadie17}. Beyond the Milky Way, studies of DM halos are restricted to using the projected kinematic properties of the stars, SCs, satellite galaxies, and gas within \citep[e.g.][]{mclaughlin99, paolillo02, peng04a, peng04b, woodley10, schuberth10, strader11,agnello14}. In the absence of gas and numerous satellites, as is the case for low-mass dwarf galaxies, a strong degeneracy between the shape of the DM density profile and the distribution of stellar orbits makes accurately mapping the distribution of DM very difficult. 

The photometric properties of SCs, on the other hand, offer a more robust method for measuring the distribution of DM within a galaxy. SCs have been observed in all types of galaxies, with population sizes ranging from over 10,000 in giant elliptical galaxies \citep{strader11} to less than 10 SCs in dwarf galaxies \citep{cole12}. A rich history of SC studies have clearly demonstrated the key role played by the host galaxy tidal field in driving the SC's dynamical evolution and the connection between small-scale properties of SCs and the large scale structural and kinematic properties of galaxies (see e.g. Heggie \& Hut 2003 as well as the recent reviews by Forbes et al. (2018) and Renaud (2018) and references therein).

Recently, \citet{contenta17} have used the mass and size of the single SC in the dwarf galaxy Eridanus II to determine whether or not the dark matter profile of Eridanus II is cuspy or cored in the inner regions. The authors use a suite of direct N-body star cluster simulations to find the initial size and mass that a progenitor SC would need to yield the present day properties of Eridanus II's central SC. Model clusters were evolved in two different dark matter density profiles that have the exact same mass within the galaxy's half-light radius, but different density profile shapes. The authors were able to reproduce the size, radial light profile, and projected position of the Eridanus II cluster in simulations where the inner dark matter density profile has a core. For a cuspy profile, the authors were only able to reproduce the observed Eridanus II cluster if its three dimensional distance is much larger than its projected distance, it has a severely inclined orbit, and the cluster just happens to be at a very specific phase in its orbit. Hence it appears that the inner dark matter profile of Eridanus II is likely cored. Similar results were found by \citet{amorisco17}, who primarily focussed on reproducing the orbital properties of the Eridanus II cluster using a collisionsless star cluster simulation.

In this study, we build upon the work of \citet{contenta17} and \citet{amorisco17} by simulating identical star clusters over a range of dark matter profiles. The purpose of our work is determine how the photometric and kinematic properties of a central SC are affected by the shape of a galaxy's density profile. 
This work is aimed at providing a theoretical framework enabling the use of SCs as tools to constrain the shape of a galaxy's density profile, effectively measuring the degree to which a profile is cored or cuspy. Furthermore, an understanding of how an underlying dark matter potential affects a SC evolution will help distinguish whether some ultra faint galaxy candidates at large galactocentric distances are dwarf galaxies embedded in a dark matter halos or star clusters devoid of dark matter \citep{conn18}. Similar to  \citet{contenta17} and \citet{amorisco17}, we will use the lone star cluster in Eridanus II as a test case. In Section \ref{s_nbody} we introduce the N-body simulations used in our study, while in Section \ref{results} we illustrate how various SC properties depend on the shape of a galaxy's dark matter density profile. We summarize our results in Section \ref{s_discussion}.

\section{N-body models} \label{s_nbody}

Each of our model clusters are simulated using the direct N-body code NBODY6TT \citep{renaud11}, which is a version of the publicly availably code NBODY6 \citep{aarseth03} that allows for clusters to evolve in an arbitrary tidal field. To mimic the tidal field of a EriII-like dwarf galaxy, we assume the dark matter density profile can be represented by a Dehnen sphere \citep{dehnen93}:

\begin{equation}\label{dsphere}
\rho(r)=\frac{M_0(3-\gamma)}{4 \pi r_0^3} (\frac{r}{r_0})^{- \gamma}(1+\frac{r}{r_0})^{\gamma-4}
\end{equation}

Where $M_0$ and $r_0$ are the scale mass and radius of the galaxy. The parameter $\gamma$ controls the shape of the galaxy's density profile, with $\gamma$=0 corresponding to a cored galaxy and $\gamma$=1 corresponding to a cuspy galaxy. For the purposes of this study, model clusters are assumed to be at the origin of the dark matter halo.

Given that the half-light radius of Eri II and the mass within this radius are $280$ pc \citep{crnojevic16} and $1.2 \times 10^7 M_\odot$ \citep{li17}, the $\gamma$=0 case corresponds to $M_0 = 4.79 \times 10^8 M_\odot$ and $r_0=877$ pc \citep{contenta17}. To explore the effects of $\gamma$ on the evolution of SCs, we also model clusters in galaxies with $\gamma$ equalling 0.2, 0.4, 0.6, 0.8 and 1.0. As per \citet{contenta17}, when $\gamma$ is changed we keep $M_0$ fixed and adjust $r_0$ so to keep the mass enclosed within the galaxy's half-light radius constant. Table \ref{table:dmparam} lists the parameters for each of the dark matter profiles that we present. It should be noted that in addition to the models listed in Table \ref{table:dmparam}, we also explored galaxy profiles where $r_0$ is kept fixed and $M_0$ is adjusted to explore whether the choice of which parameter to fix influences SC evolution. Ultimately we find that the dependence of a SC's evolution on $\gamma$, given the galaxy profiles, masses and half-mass radii explored here, do not depend on whether $r_0$ or $M_0$ is the fixed variable.

\begin{table}
  \caption{Dark Matter Density Profiles}
  \label{table:dmparam}
  \begin{center}
    \begin{tabular}{lccc}
      \hline\hline
      {$Name$} &{$M_0 (M_\odot)$} & {$r_0 (pc)$} & {$\gamma$}  \\
      \hline

{GAMMA0}& {$4.79 \times 10^8$} & {877.0} & {0.0} \\
{GAMMA02}& {$4.79 \times 10^8$} & {1000.4} & {0.2} \\
{GAMMA04}& {$4.79 \times 10^8$} & {1159.2} &{0.4} \\
{GAMMA06}& {$4.79 \times 10^8$} & {1369.6} &{0.6} \\
{GAMMA08}& {$4.79 \times 10^8$} & {1658.2} & {0.8} \\
{GAMMA10}& {$4.79 \times 10^8$} & {2071.9} & {1.0} \\

      \hline\hline
    \end{tabular}
  \end{center}
\end{table}

For each dark matter profile, we evolve a star cluster located at the center of the dark matter halo for a Hubble time. The stellar density profile of the cluster follows a Plummer model \citep{plummer11} consisting of 29,110 stars, with a mass and half-mass radius of $1.9 \times 10^4 M_\odot$ and about 10.0 pc respectively. These initial properties are such that, for the $\gamma = 0$ case, the model cluster will be comparable in size and mass to the observed SC at the center of EriII \citep{contenta17}. Within the cluster, the stellar masses are assigned assuming a Kroupa initial mass function (IMF) \citep{kroupa01} with minimum and maximum stellar masses of 0.1 and 100 $M_\odot$ respectively. Individual stellar masses will evolve following the stellar evolution prescription of \citet{hurley00} assuming a metallicity of $Z=0.001$, while any binary stars that form at later times will follow the binary stellar evolution prescription of \citet{hurley02}. Initial stellar velocities are set using a velocity dispersion calculated from the Jeans equation including the combined potential of the cluster and that of the host galaxy \citep{hernquist93}.

\section{Results} \label{results}

As we explore how a SC's evolution depends on the shape of its host galaxy's dark matter density profile, a particular focus has been placed on observable photometric and kinematic properties of the model clusters in order to maximize the applicability of our results to future studies. In the following sub-sections, we explore how the density profile, velocity dispersion and anisotropy profiles, and the degrees of energy equipartition and mass segregation within a cluster depends on the shape of the host galaxy's DM profile (as traced by $\gamma$).
 
 \subsection{Structure}

\begin{figure}
\centering
\includegraphics[width=\columnwidth]{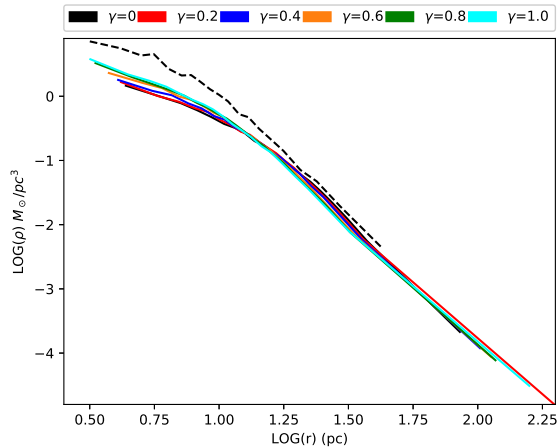}
\caption{Density profiles after 12 Gyr of star clusters in dark matter potentials with different values of $\gamma$. The initial density profile is shown as a dashed black line.}
\label{fig:rho}
\end{figure}

To illustrate the relationship between the structural evolution of a star cluster and $\gamma$, we first plot the initial and final (after 12 Gyr) three dimensional density profiles of our model clusters in Figure \ref{fig:rho}. The density profiles were determined by binning cluster stars such that $5\%$ of the total population resides in each radial bin. After 12 Gyr, clusters residing in cuspier dark matter halos are characterized by higher central densities.

The dependence of a cluster's structural evolution on the dark matter density profile can be attributed to the fact that for increasing values of $\gamma$, cluster members are confined within a deeper potential well. For an isolated cluster, the expansion triggered by mass loss due to early stellar evolution \citep[see e.g.][]{chernoff90, fukushige95, vesperini10} occurs because mass loss results in the cluster's potential decreasing. For a cluster not embedded at the centre of a DM halo potential, early mass loss due to stellar evolution can trigger a significant expansion. However, in this case, the change in the overall potential energy is minimal since the cluster resides deep within the potential well of a DM halo, minimizing the strength of the cluster expansion in response to mass loss especially in regions dominated by the DM potential. Hence, as illustrated in Figure \ref{fig:rho}, the deeper the dark matter potential (i.e. the higher the $\gamma$) the more the initial cluster's expansion is weakened. It should also be noted that, with the exception of kicked neutron stars and black holes, no stars are able to completely escape the combined potential of the star cluster and background galaxy. Hence, aside from the mass lost due to stellar evolution, the total mass of each model cluster remains constant over the course of each cluster's evolution. 

\begin{figure}
\centering
\includegraphics[width=\columnwidth]{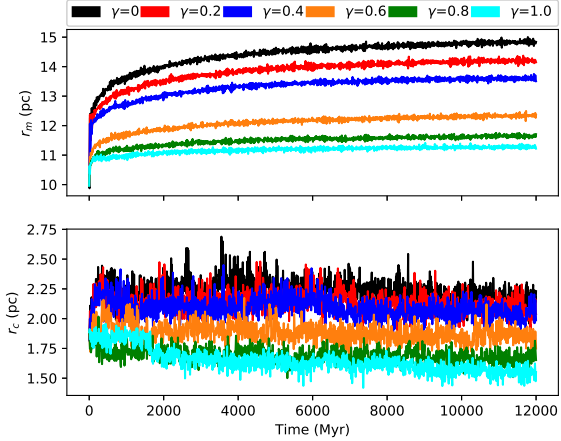}
\caption{Core (lower) and half-mass (upper) radii of star clusters in dark matter potentials with different values of $\gamma$.}
\label{fig:rm}
\end{figure}

The effects of the background potential deepening as $\gamma$ increases are also reflected in the evolution of the half-mass radius $r_m$ and the core radius $r_c$ of the cluster (see Figure \ref{fig:rm}), with cluster's in cuspier dark matter halos being less extended. The shape of the DM density profile also clearly affects any subsequent expansion of the cluster due to two-body relaxation and our simulations show that clusters in higher $\gamma$ galaxies expand less during their long-term evolution. While the differences are minor between core radii, $r_m$ can differ by up to 4 pc after 12 Gyr depending on the exact value of $\gamma$.

\subsection{Kinematics}

\begin{figure}
\centering
\includegraphics[width=\columnwidth]{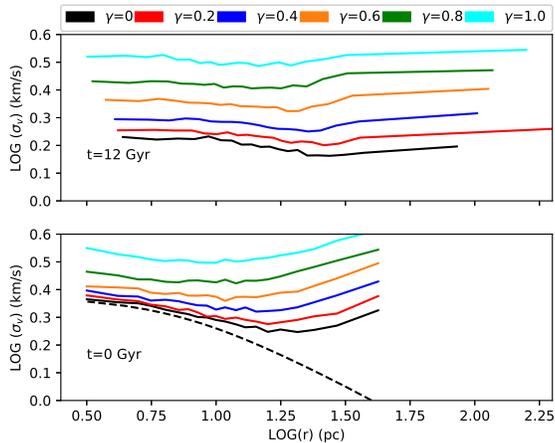}
\caption{Initial (lower) and after 12 Gyr (upper) velocity dispersion profiles of star clusters in dark matter potentials with different values of $\gamma$. For comparison purposes, the dashed black line in the lower panel illustrates the initial velocity dispersion profile of a Plummer sphere star cluster in isolation.}
\label{fig:sig}
\end{figure}

When studying the kinematic properties of model clusters, it is important to remember that these are determined not only by the cluster's own potential but are also significantly affected by the dark matter potential within which the cluster is embedded. Therefore, with respect to $\gamma$, higher $\gamma$ clusters start with a higher velocity dispersion $\sigma_v$ than lower $\gamma$ clusters due to the cuspy central background potential. The differences between the initial velocity dispersion profiles in each model cluster are illustrated in the lower panel of Figure \ref{fig:sig}, where $\sigma_v$ is the three dimensional velocity dispersion within spherical shells containing $5\%$ of all stars in the cluster. As shown in this Figure, clusters will have initial velocity dispersions greater than if they were in isolation. Additionally, it is interesting to note is that while in the cluster's inner regions the local velocity dispersion initially slightly decreases with clustercentric distance as the cluster's potential weakens, $\sigma_v$ starts to increase again as the galaxy's potential begins to dominate.

After 12 Gyr of evolution, as illustrated in the upper panel of Figure \ref{fig:sig}, the relationship between $\sigma_v$ and $\gamma$ remains, such that higher $\gamma$ clusters are characterized by a higher $\sigma_v$. However it should be noted that most of the evolution occurs early in the cluster's lifetime while its expanding due to mass loss via stellar evolution. Although the velocity dispersion is approximately constant over the entire cluster (see also \citet{yoon11}), there is still a small dip in $\sigma_v$ before a gradual increase with $r$ as the strength of the cluster's potential weakens and the galaxy's potential begins to dominate. The clustercentric distance at which $\sigma_v$ reaches a minimum is also dependent on $\gamma$, as the deeper potential well experienced by the high-$\gamma$ models results in the galaxy's tidal field becoming dominant at shorter clustercentric radii (which we will refer to as the transition radius $r_{trans}$). 

\begin{figure}
\centering
\includegraphics[width=\columnwidth]{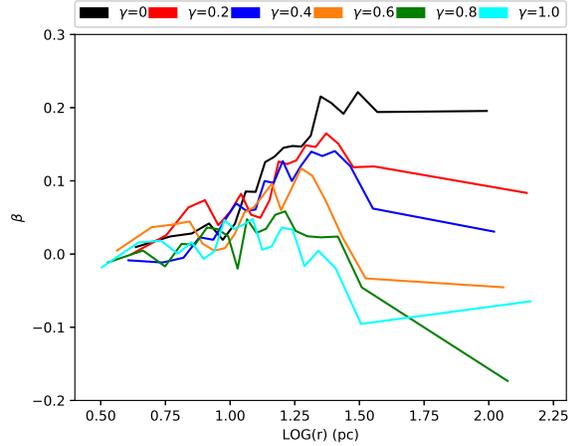}
\caption{Anisotropy profiles after 12 Gyr of star clusters in dark matter potentials with different values of $\gamma$.}
\label{fig:beta}
\end{figure}

An additional kinematic feature that we have found to have clear dependence on $\gamma$ is the anisotropy profile of the cluster, which is characterized by the radial dependence of the anisotropy parameter $\beta$:

\begin{equation}
\beta = 1 - \frac{\sigma_t^2}{2 \sigma_r^2}
\end{equation}

where $\sigma_r$ and $\sigma_t$ are the radial and tangential velocity dispersions. While all clusters start with $\beta=0$ throughout the cluster, a clear anisotropy profile quickly develops as stars lose mass due to stellar evolution at early times and the cluster expands in response to this mass loss (see Figure \ref{fig:beta}). As discussed in Section 3.1, in low-$\gamma$ models where the cluster is able to expand, orbits become preferentially radial ($\beta$ increases towards 1) with cluster centric distance out to $r_{trans}$. The orbits of stars in the high-$\gamma$ models on the other hand stay preferentially isotropic out to $r_{trans}$ as minimal cluster expansion occurs due to stellar evolution. Beyond $r_{trans}$, stellar orbits become preferentially more tangential than stars in the inner regions of the cluster. The decrease in $\beta$ beyond $r_{trans}$ can be interpreted as a consequence of the dark matter halo potential limiting the expansion of the cluster's outer regions and therefore suppressing the development of the radial anisotropy that would accompany the expansion.

\subsection{Dynamics}

With an understanding of how the structural and kinematic evolution of a cluster depends on $\gamma$, it is clear that the dynamical state of a cluster will also be affected by the shape of the host galaxy's density profile. During a cluster's long-term evolution, internal relaxation drives the cluster towards a state of equipartition via two-body interactions. Hence the dynamical state of a cluster can be measured by the relationship between stellar mass (m) and velocity dispersion, $\sigma(m) \propto m^{\eta}$, with dynamically young clusters having $\eta$ near zero and dynamical old clusters having $\eta$ values that approach -0.5 (see \citet{bianchini16} for an alternative method for measuring the degree of partial energy equipartition in a cluster). It is important to note, however, that cluster simulations have demonstrated that complete equipartition is never truly be reached \citep[e.g.][]{trenti13}.

The two-body relaxation timescale is the relevant timescale over which $\eta$ will evolve towards its maximum degree of energy equipartition. This timescale is a function of a cluster's structural and kinematical properties and varies with the distance from the cluster centre. The general expression of the two-body relaxation time $t_{relax}$ can be written as (see e.g. \citet{spitzer87}):

\begin{equation} \label{eqn:trh}
t_{relax}=\frac{0.34 \sigma_v^3}{G^2 \rho m \ln(\Lambda)}
\end{equation}

where $\rho$ is the cluster's mass density, $\sigma_v$ is the cluster's velocity dispersion, $m$ is the mean mass of stars in the cluster, and $\ln(\Lambda)$ is the Coulomb logarithm. In order to provide a general estimate of the importance of the two-body relaxation effects, the half-mass relaxation time is often used in the literature; this is an estimate of the relaxation timescale calculated adopting the average density within a cluster's half-mass radius and a global estimate of the velocity dispersion from the virial theorem. In the case of a SC embedded in a dark matter halo potential, however, one has to consider that the velocity dispersion is in part determined by the halo potential. As shown in Figure \ref{fig:sig}, the velocity dispersion of a cluster evolving at the center of a dark matter halo potential is always larger than that of an isolated cluster with the same density profile. This implies that the relaxation timescale of a cluster in a dark matter halo potential is longer than the corresponding isolated system, consistent with the findings of \citet{yoon11}. For the systems we have explored here the cluster's velocity dispersion increases with $\gamma$ (as shown in Figure \ref{fig:sig}) and the average density within the cluster half-mass radius decreases with $\gamma$ (see Figure \ref{fig:rho}), but it is the velocity dispersion trend that is dominant as relaxation timescales increase with $\gamma$ from 2900 Myr ($\gamma=0$) to 13,800 Myr ($\gamma=1$).

\begin{figure}
\centering
\includegraphics[width=\columnwidth]{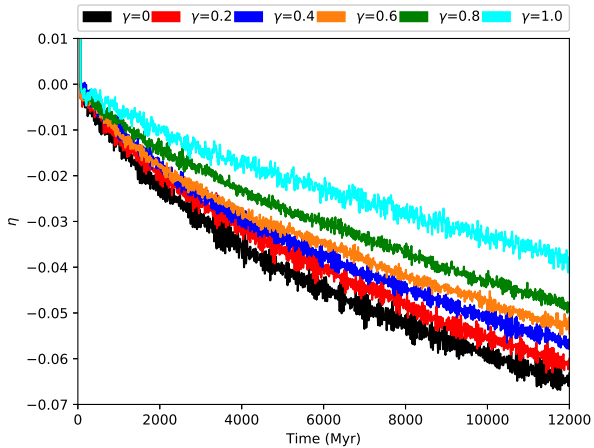}
\caption{The evolution of $\eta$ for star clusters in dark matter potentials with different values of $\gamma$.}
\label{fig:eta}
\end{figure}

%For clusters with the same mass loss history, which is the case for our models, it will be differences in their velocity dispersions and densities that set their relative relaxation times. 
The evolution of $\eta$ shown in Figure \ref{fig:eta} reflects the differences in the two-body relaxation timescales of clusters in different dark matter halos: clusters evolving in low-$\gamma$ halos relax at a faster rate (and $\eta$ undergoes a more rapid evolution) than those embedded in  high-$\gamma$ halos. Hence the high $\sigma_v$ associated with high-$\gamma$ values results in them having longer relaxation times and being dynamically younger than low-$\gamma$ models. 

\begin{figure}
\centering
\includegraphics[width=\columnwidth]{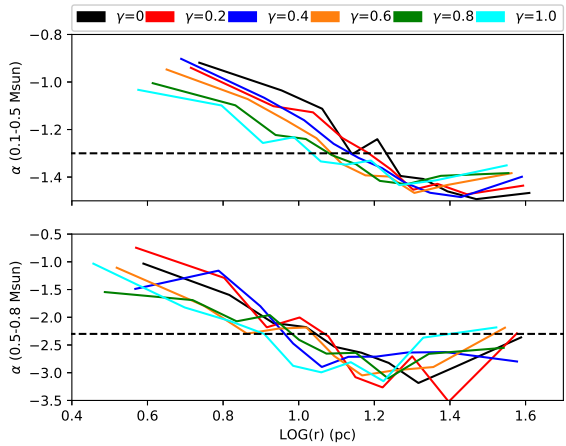}
\caption{Radial variation in the slope of the mass function between $0.1-0.5 M_\odot$ (upper panel) and $0.5-0.8 M_\odot$ (lower panel) $\alpha$ after 12 Gyr of star clusters in dark matter potentials with different values of $\gamma$. The horizontal dashed line illustrates the initial value of $\alpha$ over the 0.1-0.5 $M_\odot$ mass range.}
\label{fig:alpha}
\end{figure}

With an observational measurement of $\eta$ being quite difficult, the dynamical state of a cluster can alternatively be probed by measuring the variation of the slope of the stellar mass function $\alpha$ with radial distance from the cluster centre. As a cluster relaxes, repeated two-body interactions result in low-mass stars gaining kinetic energy and migrating outwards while high-mass stars lose kinetic energy and fall towards the centre of the cluster. Over time, the inner mass function of a cluster will become dominated by high-mass stars while the outer mass function will become dominated by low-mass stars. As illustrated over the $0.1-0.5 M_\odot$ and $0.5-0.8 M_\odot$ mass ranges in Figure \ref{fig:alpha}, after 12 Gyr the slope of the stellar mass function evolves from its initial value (horizontal dashed line) by various degrees in each cluster. 

Figure \ref{fig:alpha} illustrates that, for stars between $0.1-0.5 M_\odot$, model clusters in low-$\gamma$ halos develop a stronger radial variation in $\alpha$ than clusters in high-$\gamma$ halos. However, over the higher mass range ($0.5-0.8 M_\odot$) the trend is  weaker. This difference is due to the fact that the local relaxation time varies with clustercentric distance, such that the outer regions are characterized by very long relaxation timescales and therefore essentially do not contribute to the segregation process. Hence radial variation in the slope of the mass function is primarily due to the outward migration of low-mass stars from the cluster's inner regions, where the local relaxation time is shorter, and some inward migration of massive stars that initially populate the cluster's inner and intermediate regions. With respect to the radial profiles of $\alpha$ in Figure \ref{fig:alpha}, a typical variation in $\alpha$ is observed in the inner regions due to the effects of relaxation. In the outer regions, the profiles flatten at values of $\alpha$ corresponding to MFs that are only slightly steeper than the IMF (in particular for the cuspier DM halo cases).

\begin{figure}
\centering
\includegraphics[width=\columnwidth]{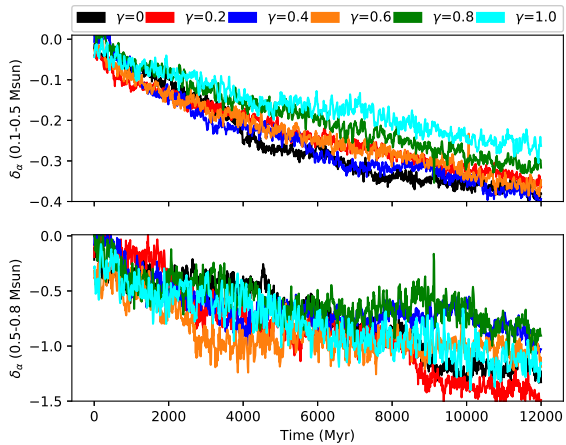}
\caption{The evolution of $\delta_\alpha$ for stars between $0.1-0.5 M_\odot$ (upper panel) and $0.5-0.8 M_\odot$ (lower panel) for star clusters in dark matter potentials with different values of $\gamma$.}
\label{fig:da}
\end{figure}

The evolution of the radial variation in the slope of the stellar mass function, which provides a method for measuring the degree of mass segregation (i.e. the dynamical state) of a cluster, can be quantified using $\delta_\alpha = d\alpha(r)/d(\ln(r/r_m))$ \citep{webb16}. The complete evolution of $\delta_\alpha$ for each model cluster using the $0.1-0.5 M_\odot$ and $0.5-0.8 M_\odot$ mass ranges is illustrated in Figure \ref{fig:da}. It should be noted that we have only measured $\delta_\alpha$ out to 1.6 $r_m$ , within the $r_{trans}$ of the highest $\gamma$ halo, in order to avoid the flattening and increase in the $\alpha(r)$ profile beyond $r_{trans}$ affecting our measurement.

The evolution of $\delta_\alpha$ mimics that of $\eta$, consistent with the results of \citet{webb17}, since neither of these models clusters undergo core collapse \citep{bianchini18}. Low-$\gamma$ clusters are able to reach a higher degree of mass segregation than high-$\gamma$ clusters in agreement with the expectations based on the dependence of the relaxation timescale and the strength of background tidal field on $\gamma$ discussed above. Therefore clusters evolving at the centre of cuspy dark matter halos will be dynamically younger than clusters in cored dark matter halos.
 
As pointed out in Section 3.1, with the exception of ejected neutron stars and black holes, the clusters do not lose stars during their evolution. We emphasize that, despite the model SCs showing some evidence of mass segregation, the global mass function is not altered during the cluster's evolution (aside from the high-mass end changing due to stellar evolution). Alternatively, for example, the tidal stripping of mass segregated globular clusters in extensive tidal fields results in the preferential escape of low-mass stars and the gradual flattening of the low-mass side of the mass function. Therefore in stellar systems evolving at the center of dark matter halos, like SCs and ultrafaint-dwarf galaxies, the present-day mass function will be the same as the initial mass function; variations in the observed global present-day mass function such as those reported by \citet{geha13} and \citet{gennaro18} would, as pointed out by those authors, imply a non-universal initial mass function. 

%M/L Ratio of 100-600 for the GCs, M*/L* ratio ~2
 
\section{Discussion and Summary}\label{s_discussion}

In this paper we have explored the structural and kinematic evolution of a star cluster evolving at the centre of the dark matter halo of a dwarf galaxy. We have simulated the evolution of initially identical model star clusters at the centre of six different dark matter halos. The choice of initial conditions adopted in this paper is informed by the properties of the ultrafaint dwarf Eri II and its central star cluster. Hence the mass enclosed within the current half-light radius of EriII has been kept fixed for each dark matter halo, while we have varied the shape of the dark matter halo density profile by exploring a range of values of the parameter $\gamma$ (see Equation 1) including cuspy and cored halos.

The main processes driving the evolution of our clusters are mass loss due to stellar evolution which triggers an early cluster's expansion and two-body relaxation driving the cluster's long-term evolution. Our simulations indicate that star clusters evolving in cored dark matter halos undergo a stronger early expansion and are characterized by lower central densities, larger half-mass radii and lower velocity dispersions. The early cluster expansion also leads to the development of a strong radial anisotropy in the velocity distribution that increases with clustercentric distance (see Figure \ref{fig:beta}). As $\gamma$ increases, so will the cluster's central density and velocity dispersion, while the half-mass radius decreases. In dark matter halos with larger values of $\gamma$, the stronger external field significantly suppresses early cluster expansion which results in systems characterized by weaker radial anisotropy or isotropy in the velocity distribution (see Figure \ref{fig:beta}). The increase of $\sigma_v$ with $\gamma$ is also of particular importance, as it implies that clusters in higher $\gamma$ halos have longer relaxation times.  Hence clusters in higher $\gamma$ halos are dynamically younger than clusters in lower $\gamma$ halos and are characterized by a lower degree of energy equipartition (see Figure \ref{fig:eta}) and mass segregation (see Figures \ref{fig:alpha} and \ref{fig:da}). 

While measurements of the degree of energy equipartition and mass segregation in cluster's at distances comparable Eridanus II are likely not currently possible, future observational studies (especially with the upcoming James Webb Space Telescope) may make such measurements a reality. Furthermore, additional simulations of star clusters with a range of initial properties embedded in a range of dark matter halos will allow for the density profiles of dwarf galaxies to be accurately measured and constraints to be placed on the role and nature of dark matter in the Universe. Preliminary work suggests that denser clusters, which will be less affected by a central dark matter halo, are able to expand, segregate at a faster rate due to shorter relaxation times, and develop steep radial anisotropy profiles regardless of $\gamma$. Extended clusters on the other hand, with stars near or even beyond $r_{trans}$, will undergo very little structural or kinematic evolution (even less than the model cluster in the $\gamma=1$ halo presented here).

Between the studies of  \citet{contenta17},  \citet{amorisco17}, and the results presented here, it is becoming abundantly clear that star clusters in dwarf galaxies offer a unique window into the field of cosmology and that their structural and kinematic properties can provide a significant insight into a number of fundamental issues concerning the structure of dark matter halos. Furthermore, the results presented in this work concerning the specific fingerprints of a host dark matter halo on a central star cluster can also help to determine whether a stellar system is an ultra diffuse galaxy or a star cluster based on whether or not the population appears to be under the influence of a background dark matter potential. 

\section*{Acknowledgements}
JW acknowledges financial support through research grants and scholarships from the Natural Sciences and Engineering Research Council of Canada. This work was made possible in part by Lilly Endowment, Inc., through its support for the Indiana University Pervasive Technology Institute, and in part by the Indiana METACyt Initiative. The Indiana METACyt Initiative at IU is also supported in part by Lilly Endowment, Inc.

\bsp

\label{lastpage}

\end{document}